\documentclass[prx,aps, twocolumn, showpacs,groupedaddress,10pt,nofootinbib]{revtex4-2} 

\usepackage[T1]{fontenc}
\usepackage{multirow}
\usepackage{bm,dcolumn}
\usepackage{amsmath,amsfonts,amssymb,amsthm,amscd} 
\usepackage{graphicx}
\usepackage{verbatim}
\usepackage[section]{placeins}
\usepackage{color}
\usepackage{soul}
\usepackage{xparse}
\usepackage{physics}
\usepackage{siunitx}
\usepackage[colorlinks,linkcolor=blue,citecolor=blue,anchorcolor=blue]{hyperref}

\graphicspath{ {./Figures/} }
\usepackage{pdfpages} 
\usepackage{pgffor}

\makeatletter
\AtBeginDocument{\let\LS@rot\@undefined}
\makeatother

\begin{document}

\title{High-fidelity entanglement between a telecom photon and a room-temperature quantum memory}

\author{Yang Wang}
\author{Alexander N. Craddock}
\author{Jaeda M. Mendoza}
\author{Rourke Sekelsky}
\author{Mael Flament}
\author{Mehdi Namazi}
\email{mehdi@quconn.com}
\affiliation{Qunnect Inc. 63 Flushing Ave, Bldg. 303, Ste. 701, Brooklyn, NY 11205-1005}
\date{\today}

\begin{abstract}
Entanglement distribution through existing telecommunication infrastructure is crucial for realizing large-scale quantum networks. However, distance limitations imposed by photon losses and the no-cloning theorem present significant challenges. Quantum repeaters based on entangled telecom-wavelength photons and quantum memories offer a promising solution to overcome these limitations. In this work, we report an important milestone in quantum repeater architecture by demonstrating entanglement between a telecom-wavelength (1324 nm) photon and a room-temperature quantum memory with a fidelity up to 90.2\%, using simple rubidium systems for both photon generation and storage. Furthermore, we achieve high-rate photon-memory entanglement generation of up to 1,200 Bell pairs per second with 80\% fidelity. The technical simplicity and robustness of our room-temperature systems paves the way towards deploying quantum networks at scale in realistic settings.

\end{abstract}

\maketitle

\section{Introduction}

Quantum networks \cite{Kimble2008,Awschalom2021}, with their foundation of distributed entanglement, are crucial for a wide range of transformative applications, including connecting quantum computers \cite{Cirac1999,Cacciapuoti2020}, secure quantum communication \cite{Gisin2002,Ekert2014,Scarani2009,Xu2020}, and enhancing quantum sensing capabilities \cite{Komar2014,Guo2020}. Establishing a practical quantum network over existing telecommunication infrastructure often necessitates the paradigmatic quantum repeater protocol \cite{Briegel1998,Duan2001,Sangouard2011,Beukers2024}, which integrates stationary qubits (e.g., quantum memories) and flying qubits (e.g., photons). 
The past decade has witnessed tremendous progress towards such quantum repeaters on various physical platforms such as trapped ions \cite{Krutyanskiy2024,Saha2024}, cold atoms \cite{Zhou2024,Liu2024}, rare-earth-doped crystals \cite{Rakonjac2021}, and defects in diamonds \cite{Hermans2022,Knaut2024}. High rate \cite{Stephenson2020}, high fidelity \cite{Krutyanskiy2024}, long distance \cite{Leent2020}, and memory enhanced \cite{Bhaskar2020} quantum communication have been demonstrated.

Implementing quantum networks in practical environments presents several challenges, including telecom wavelength compatibility, space and power constraints in fiber hubs and data centers, and requirements for robustness, scalability, and usability. Warm vapor technology emerges as a promising solution that addresses both the scientific requirements and hardware demands of quantum networks. On the former, researches have demonstrated high-performance photon sources \cite{Park2019,Davidson2021,Chen2022,source2024} and quantum memories with long storage time \cite{Katz2018, Dideriksen2021}, high bandwidth \cite{Wolters2017,Buser2022}, high fidelity\cite{FLAME,ORCA,wang2022}, and high efficiency \cite{Ma2021}, enabling novel applications such as photon synchronization \cite{Davidson2023}. On the latter, 
these vapor-based technologies operate without the need for complex cryogenic environments or laser-cooling systems typically required by other approaches, resulting in systems that are robust, scalable, and field-deployable. We have previously developed \cite{wang2022,source2024} and deployed \cite{APC2024} warm-vapor-based devices in a networking testbed at a metropolitan scale.

\begin{figure}
\includegraphics[width=8.6cm]{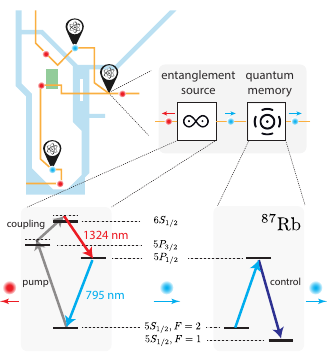}
\caption{A quantum network based on warm-vapor devices. Each node comprises an entanglement source and a quantum memory, both using warm rubidium ($^{87}$Rb) vapors with identical atomic levels for native compatibility. The source is based on off-resonant four-wave-mixing (FWM) \cite{source2024}, producing entangled photon pairs with one in the telecom wavelength (1324 nm, shown in red) and one in the near-infrared (NIR) wavelength (795 nm, shown in blue). The memory, utilizing electromagnetic-induced transparency (EIT) \cite{wang2022}, stores the NIR photon in the atomic ground state. Entanglement is distributed by the traveling telecom photons (red balls between nodes) and entanglement swapping (not shown), while the rate is enhanced by the memory storing the NIR photon (blue balls at the nodes). 
}
\label{fig:1}
\end{figure}

Here, we present an architecture for quantum networks based on room-temperature rubidium ($^{87}$Rb) vapor devices and demonstrate an important milestone in this approach: storing photons from an entanglement source in a quantum memory. Our architecture is sketched in Fig. \ref{fig:1}. 
At each node, a pair of entangled photons is generated by an entanglement source. One photon from the pair is stored in a quantum memory and the other is distributed across the network. Distant nodes are entangled recursively via entanglement swapping where the rate is greatly enhanced by the quantum memory. Our approach leverages identical atomic systems for both entangled photon pair generation \cite{source2024} and photon storage \cite{wang2022}, directly establishing entanglement between a telecom photon (1324 nm) and a quantum memory without additional resources such as quantum frequency conversion (QFC). This design achieves native compatibility and field deployability while maintaining optimal performance. Unlike merged source-memory schemes \cite{Duan2001}, our separated architecture optimizes critical physical parameters such as optical density, beam sizes, and power requirements \cite{wang2022, source2024}. The technical simplicity of our solution yields high entanglement generation rates---a critical performance metric for quantum network protocols.

We demonstrate polarization entanglement between a telecom photon and a quantum memory with fidelity up to 90.2\%. This performance approaches state-of-the-art systems such as cold atoms (89.7\% with QFC \cite{Leent2020}), NV centers (77\% with QFC \cite{Tchebotareva2019}), rare-earth doped crystals (86\% \cite{Jelena2022}), and trapped ions (96\% with QFC) \cite{Krutyanskiy2024}), while eliminating the need for laser cooling or cryogenic environments. Furthermore, we achieve a high photon-memory entanglement rate exceeding 1,200 pairs per second, while maintaining a fidelity of 80\% (equivalent of Clauser-Horne-Shimony-Holt \cite{Clauser1969} $S>2$). Additionally, our memory exhibits high bandwidth (capable of storing photons up to $2\pi\times$266 MHz) and high, source independent, signal-to-noise ratio $\rm{SNR}_{\langle n \rangle=1}=95(5)$, facilitating compatibility with alternative photon sources including quantum dots \cite{Tomm2021} and narrow band spontaneous parametric down-conversion (SPDC) sources \cite{Mottola2020}. 
These advancements build upon our previous work \cite{wang2022, source2024, APC2024} and establish a foundation for real-world quantum network deployment.

The structure of this paper is as follows: Sec. \ref{sec:exp} presents our experimental methodology and apparatus; Sec. \ref{sec:memory} demonstrates high-performance memory results with both weak coherent pulses and single photons from our entanglement source---a critical prerequisite for photon-memory entanglement; Sec. \ref{sec:entanglement} benchmarks entanglement fidelity between a quantum memory and a telecom photon; and Sec. \ref{sec:extra} analyzes memory utility time and entanglement generation rates.

\section{Experimental setup}
\label{sec:exp}

\begin{figure*}[t!]
\includegraphics[width=18.2cm]{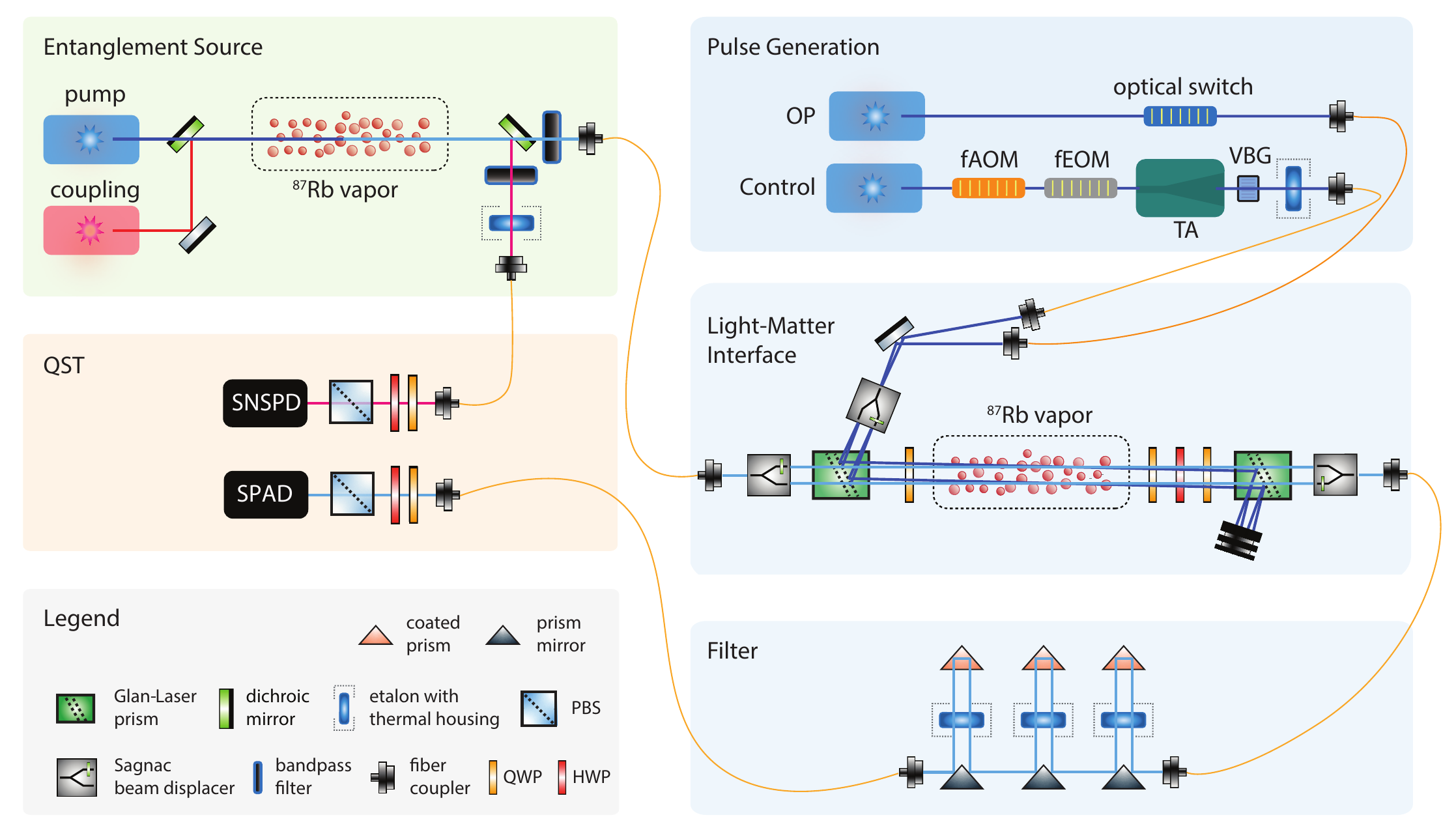}
\caption{
Simplified schematics of the experimental setup. The entanglement source (detailed in Ref \cite{source2024}) uses hot Rb vapor optically excited by a pump and a coupling field. The generated photon pairs are separated by a dichroic mirror and filtered by band-pass filters to remove the strong pump and coupling fields. The telecom photon (red) is further filtered by a thermally stabilized optical cavity to obtain a single spectral mode, while the NIR photon is routed to the quantum memory.
The quantum memory consists of three modules: pulse generation, light-matter interface, and filter. The pulse generation module creates a strong, high-bandwidth, high-extinction control pulse as well as a state-preparing optical pumping (OP) field. The light-matter interface module stores and retrieves input photons (e.g., from an entanglement source), similar to Ref \cite{wang2022}, includes an extra OP field intersecting the Rb atoms at a small angle. The filter module has three temperature-stabilized cavities each arranged in a double-pass fashion to remove noise photons due to control field. Each cavity is slightly titled to avoid interference from back reflections. The memory retrieved NIR photon and the source telecom photon are measured at the quantum state tomography (QST) station. Abbreviations: SNSPD, superconducting nanowire single-photon detector;
SPAD, single-photon avalanche diode;
PBS, polarization beamsplitter;
HWP, half waveplate;
QWP, quarter waveplate;
fEOM, fiber electro-optic modulator;
fAOM, fiber acoustic-optical modulator;
TA, tapered amplifier;
VBG, volume Bragg grating.
}
\label{fig:2}
\end{figure*}

Figure \ref{fig:2} shows the simplified experimental setup, including the source, the memory, and the entanglement analyzer. The memory has three functional blocks: pulse generation, light-matter interface, and filter. For simplicity, control and measurement electronics are not shown. 

The source is based on four-wave-mixing (FWM) in a warm Rb vapor \cite{Willis2010,source2024}, producing a pair of a telecom photon and a near-infrared (NIR) photon (795 nm) in one of the Bell states: $|\Phi^+\rangle=\frac{1}{\sqrt{2}}(|HH\rangle+|VV\rangle)$, and $H(V)$ is horizontal (vertical) polarization. The physical setup of the source (Fig. \ref{fig:2}, upper left) is identical to the one described in Ref \cite{source2024} with only one difference: the telecom photons are filtered by an optical cavity [full width at half maximum (FWHM): 266(5) MHz; free spectral range (FSR): 15.2(4) GHz] before being routed to an entanglement analyzer via a single mode fiber. The pump field (780 nm) is locked 1.1 GHz blue-detuned from the $|5S_{1/2}, F=2\rangle \rightarrow |5P_{3/2}, F'=3\rangle$ transition, while the coupling field (1367 nm) is locked 3.5 GHz red-detuned from the $|5P_{3/2}, F=3\rangle \rightarrow |6S_{1/2}, F'=2\rangle$ transition. 
We typically run the source with a 10 mW coupling field and a 1 mW pump field. The Rb vapor is temperature-stabilized to $ 95(1) ^{\circ}C$. The cavity resonance is set to +700 MHz relative to the telecom photon decaying to $|5P_{1/2}, F=1\rangle$ state. In other words, detection of such a telecom photon heralds an NIR photon that is -700 MHz detuned from $|5P_{1/2}, F=1\rangle \rightarrow |5S_{1/2}, F'=2\rangle$ transition. This resonance position is optimized for good memory storage efficiency \cite{Buser2022}, high source heralding efficiency ($\eta$), and high source rate. See the Appendix for details.

The memory is based on electromagnetically induced transparency (EIT) \cite{Fleischhauer2000,Eisaman2005,Gorshkov2007} in a $\Lambda-$system on Rb atoms (see level schemes in Fig. \ref{fig:1}). A strong control field $\Omega_c$ coherently maps the NIR photon to and from a collective excitation of Rb atoms in the hyperfine ground state, realizing photon storage and retrieval, respectively. The control field is $\sigma^+$ polarized while the NIR photon is $\sigma^-$ polarized. This arrangement leverages the polarization selection rules and eliminates the noise from the FWM process (note it is different from the source FWM), see \cite{Lauk2013,WALTHER2007,Zhang2014}. The $\Omega_c$ field is frequency stabilized to be detuned -700 MHz from the $|5S_{1/2}, F=1\rangle \rightarrow |5P_{1/2},F'=1\rangle$ transition to match the source NIR photon. This large detuning, however, would lead to insufficient state preparation, leaving some atoms in the $F=1$ state. 
These $F=1$ atoms would then cause noise photons during the memory retrieval due to spontaneous Raman scattering \cite{Raymer1985}, and lower the retrieval fidelity \cite{wang2022}. To solve this problem, we use a second laser field resonant with the $|5S_{1/2}, F=1\rangle \rightarrow |5P_{1/2},F'=2\rangle$ transition to provide sufficient optical pumping (OP). This OP field is turned off during the storage. 

The operating bandwidth of the memory is primarily determined by the strength and the bandwidth of the $\Omega_c$ field \cite{Wolters2017,Wei2020,Buser2022,Shinbrough2023}. To store and retrieve photons of GHz-level bandwidth, one needs to generate a control field pulse with high bandwidth (sub-nanosecond rise/fall time) and high Rabi frequency ($\sim$ 1 W for mm-sized beams). 
Pulses with these characteristics, unfortunately, fall through the gaps of existing commercial technologies. With CW-lasers, light modulators that can directly handle high power (usually free-space ones) tend to have a lower bandwidth (e.g., acoustic optical modulators, AOMs). With nanosecond pulsed lasers, getting on-demand pulses with the laser frequency stabilized to atomic references is challenging. 
One very promising method is to first generate high-bandwidth pulses at low optical power, e.g., using a fiber electro-optic modulator (fEOM), and then amplify such pulses to high optical power, e.g., using a tapered-amplifier (TA). This approach has been studied by many groups \cite{Rogers2016, Kaufman2017, He2020, Clarke2022}, and has been successfully applied to high-bandwidth quantum memory \cite{Wolters2017,Buser2022, Mottola2023}. Our pulse generation is based on this approach, see Fig. \ref{fig:2} upper right. We first generate a weak pulse (2 mW peak power) of 5 ns with 0.3 ns fall-time using an fEOM 
driven by a digital delay pulse generator 
A fiber AOM 
with a high power extinction ratio (60 dB) is installed to overcome the low power extinction of the fEOM (35 dB), which causes unintentional retrieval and lowers the memory storage time substantially. 
The resulting pulse is amplified by a TA 
to $\sim $2 W, and spectrally filtered by a volume-Bragg grating (VBG) and a monolithic optical cavity [0.55(4) GHz FWHM, 30.5(4) GHz FSR] to remove the excessive broadband noise photons from TA due to the amplified spontaneous emission (ASE).

The light-matter interface (Fig. \ref{fig:2}, center right) stores and retrieves the NIR photons in and from a warm Rb vapor. To preserve the quantum state of the NIR photons, we separate the polarization components into two spatial modes (rails) for memory operation and recombine them afterwards. This is accomplished using Sagnac-like beam displacers \cite{Salazar2015,wang2022}. The (dual-rail) source photon and the (dual-rail) control field ($\Omega_c$) are combined with a Glan-Laser (GL) prism and interact with a warm vapor of $^{87}$Rb (99\% purity) at 57(1)$^{\circ}$C in a glass cell with 2 Torr Ne buffer gas, housed in a two-layer magnetic shield (not shown). 
After separating the $\Omega_c$ field with a second GL, the source photon is converted back to a single spatial mode with polarization encoding and coupled into a fiber for further processing. The $1/e^2$ beam diameters are 1.1 mm for the control light and 0.93 mm for the source photon. With a typical control power of 410 mW per rail, the peak Rabi frequency for the control field reaches $2\pi \times 152$ MHz. 
The dual-rail OP field co-propagates with $\Omega_c$ at a small angle (8.5 mrad) to pump atoms to $|5S_{1/2},F=2\rangle$ state. The total transmission of this light-matter interface module is $66(1)\%$.

The storage of the source NIR photon effectively entangles the telecom photon with the quantum memory. 
Using the memory in specific quantum repeating protocols, however, requires retrieving the NIR photon, where the readout noise becomes the dominant error in the fidelity. 
Most spectral filtering methods, like high-finesse optical cavities \cite{wang2022}, achieve noise suppression at the cost of lower linewidth due to the Lorentzian line shape. This bandwidth mismatch in turn reduces the transmission of the (high-bandwidth) memory-retrieved photon through the filter and lowers the entanglement rate. We solve this challenge with a high-bandwidth filter design, see Fig. \ref{fig:2} (lower right). Employing a multi-pass optical setup with low-finesse, high-linewidth optical cavities [1.55(6) GHz bandwidth, 60.2(5) GHz FSR], we modify the Lorentzian function to achieve the desired noise suppression ($\sim114$ dB) while maintaining high overall bandwidth ($2\pi\times182$ MHz). The total transmission of this filtering module is $35(1)\%$.

Entanglement fidelity is evaluated using standard quantum state tomography (QST) \cite{James2001} (Fig. \ref{fig:2} center left) between the source telecom photon, measured with a superconducting nanowire single-photon detector (SNSPD, 90\% quantum efficiency, 94-ps timing jitter), and the memory-retrieved NIR photon, measured with a single-photon
avalanche diode (SPAD, 65\% quantum efficiency, 350-ps timing jitter). The transmission of this QST module is typically 90(1)\%. 

For characterizing the photon-memory entanglement we synchronize the memory operation to the source. The source continuously produces entangled photon pairs. Upon detecting a telecom photon, we trigger the memory sequence to store and retrieve the NIR photon, similar to \cite{Buser2022}. 
In our setup, the detection of a telecom photon triggers a digital delay generator (DDG), which drives the fEOM and fAOM to create the $\Omega_c$ field sequence. It also controls the OP field to prepare Rb atoms for storage. We timestamp the DDG event (corresponding to a telecom photon detection) and the NIR photon detection at the QST using a time-tagger with 1-ps resolution to analyze coincidences. To compensate for the 25-meter separation between the source and the SNSPD, as well as the internal latencies of various electronic devices (e.g., SNSPD and DDG), we buffer the NIR photon with a 100-meter fiber before reaching the memory.

\section{high-performance quantum memory}
\label{sec:memory}

A high-performing quantum memory is critical for memory-photon entanglement. Here we study the performance of the quantum memory in two ways. First, we demonstrate the storage and retrieval of a weak-coherent-state (WCS) \cite{Beukers2024} pulse with a bandwidth similar to our entanglement source, verifying the memory's high bandwidth and high signal-to-noise ratio (SNR) in isolation. Second, we present results on the memory operations with single photons from our entanglement source. The consistent high performance across both measurements establishes our quantum memory as a critical component for effective entanglement distribution.

\subsubsection{Memory solo operation}
In the first experiment, we operate the memory to store and retrieve a synthetic pulse created from a laser source containing $\langle n \rangle < 1$ photon, i.e., a WCS pulse. To mimic the photon of our entanglement source, we perform high-bandwidth amplitude modulation with an fEOM to a laser field where its frequency is stabilized to be -700 MHz from the $|5P_{1/2}, F=1\rangle \rightarrow |5S_{1/2}, F'=2\rangle$ transition (see Sec. \ref{sec:exp}). The fEOM is driven by an arbitrary waveform generator (AWG) with 1 ns resolution. The generated pulse has a Gaussian shape and a FWHM bandwidth of $2\pi\times 142(4)$ MHz. This bandwidth, still lower than our entanglement source, is mainly limited by the bandwidth of the AWG. 

\begin{figure}[t!]
\includegraphics[width=8.6cm]{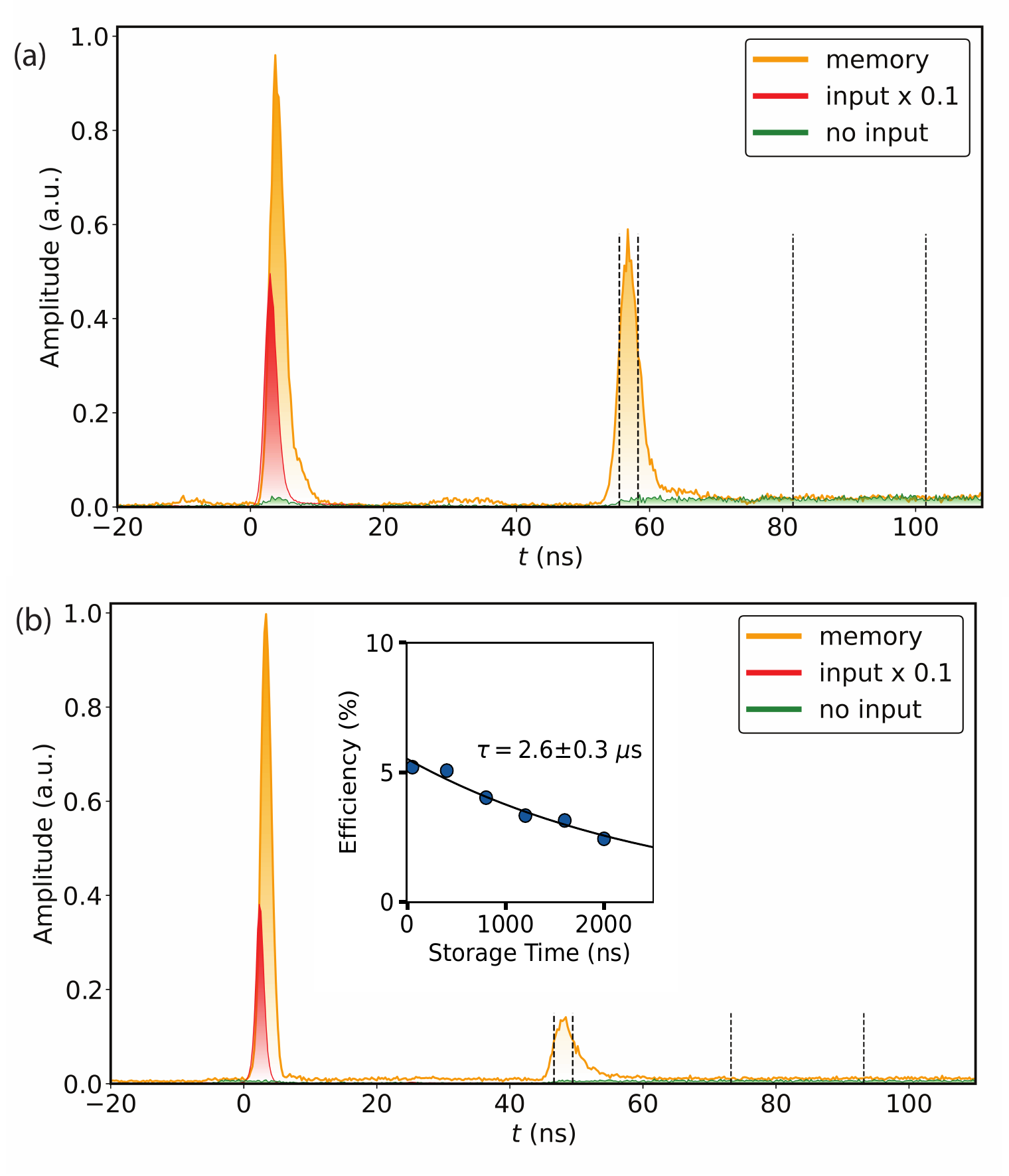}
\caption{High-performance memory operation for a weak coherent state (WCS) pulse (a) and single photons (b) from the entanglement source. In both cases, we plot the coincidence histogram under three conditions: {\bf{memory}} (orange), where a photon/pulse is stored and later retrieved; {\bf{input}} (red), where Rb vapor is removed and both control and OP fields are blocked; {\bf{no input}} (green), where only the input photon/pulse is blocked.
The dashed lines mark the different detection windows for analyzing signal and noise. 
Insert: memory coherence time. 
}
\label{fig:3}
\end{figure}

We synchronize the memory sequence and the WCS generation to a global clock at 500 kHz, and define the arrival of the WCS pulse at the Rb vapor as $t=0$. The main steps of memory operation are as follows (see Fig. \ref{fig:3}a): first, the optical pumping is turned off at $t=-20$ ns; then, a 5-ns (0.3 ns fall-time) control pulse $\Omega_c$ with 410 mW (per rail) peak power reaches the Rb vapor, with its falling edge aligned to coincide with the WCS pulse (t$\sim$0 ns); at $t=55$ ns, $\Omega_c$ field with the same power is pulsed again to retrieve the photon and remains on until the next cycle. The OP field is turned back on at $t=150$ ns.

In Fig. \ref{fig:3}a we plot the measured histogram of coincidences under three experimental conditions: (i) memory operation (orange trace), where the WCS pulses are stored and retrieved; (ii) input (red trace, scaled by a factor of 0.1), with $\Omega_c$ and OP field blocked, and memory Rb vapor removed; (iii) no input (green trace), where the WCS pulses are blocked. The histograms show coincidences between the SPAD events and the cycle trigger accumulated over 2 seconds with a 256-ps bin size. To analyze the memory SNR, we choose a detection window (dashed lines) of 2.82-ns, integrate the coincidences under the memory data (orange) to get the signal. We then select a larger window (dashed lines) of 20-ns taken far after the photon retrieval (where $\Omega_c$ remains on), to determine the noise (by integrating coincidences under the memory data) with better photon statistics, which is consistent with the no input case (green trace). The choice of the 2.82-ns detection window size is a trade-off between the SNR and the number of successful events that includes at least 50\% of the retrieved photon, see Sec. \ref{sec:extra} and Fig. \ref{fig:6} for a detailed discussion. 
With this choice, we obtain an SNR of 31(2), an internal storage efficiency of 9.5(5)\% (ratio of the retrieved photons to the input photons through the memory), and an unconditional noise floor of $1.2\times10^{-4}$ per trial.

The SNR measured above scales with the input photon number, and a more useful metric to characterize the memory solo performance is the SNR for an input pulse with $\langle n\rangle=1$ photon \cite{Jobez2015}: $\rm{SNR}_{\langle n \rangle =1}$, which allows one to compare the memory performances in isolation across different technical implementations. The photon number of the input pulse is determined by integrating the coincidences under the input data (Fig. \ref{fig:3}a, red trace), and backtracked to take into account of the transmission of the memory system and the SPAD efficiency. We obtain $\langle n\rangle = 0.32 (2)$. The normalized SNR is then $\rm{SNR}_{\langle n \rangle =1} = 95(5) $. This high SNR marks nearly 5 times improvement over our past work \cite{wang2022}, which we attribute to the fast retrieval. 
We note that this observation has also been reported by other groups \cite{Buser2022}.    

\subsubsection{Memory source operation}

Having demonstrated high performance of memory in isolation, we proceed to integrating our memory with single photons generated by our entanglement source. Different from the memory solo operation, we synchronize the memory sequence to the entanglement source. 
Specifically, the source continuously produces photon pairs while the memory operates only upon detection of a telecom photon. This approach ensures optimal data rates without requiring source pulsing ---a methodology widely adopted in the field for characterization \cite{Willis2010,Buser2022}. We emphasize that while this arrangement sufficiently demonstrates proof-of-concept capabilities, it does not preclude alternative operational modes, such as globally clocked mode in the previous memory solo experiments, which more closely represents real-world implementation scenarios.

Our entanglement source produces entangled photon pairs in the Bell state $|\Psi^+\rangle$. For the purpose of characterizing the memory performance, we only analyze photon pairs in a fixed polarization state, $|HH\rangle$. 
In the next section, we will analyze a set of polarization states to get a comprehensive understanding of the entanglement state between the memory and the telecom photon.

The memory source operation is identical to the memory solo operation with the cycle trigger replaced by the detection of a telecom photon. Running the source with a 10-mW coupling and 1-mW pump we obtained a telecom rate of $2.1\times10^5$ count per second after passing through the optical cavity with an FWHM of 266(5) MHz. Fig. \ref{fig:3}b shows the experimental results, where the histograms are of coincidences between the NIR and the telecom photons accumulated over 20 seconds with a 256-ps bin size. For a 2.82-ns detection window, we obtain an SNR of 9.8(3), an internal storage efficiency of 5.2(1)\%, and an unconditional noise floor of $8.8\times10^{-5}$ per trial.

Fig. \ref{fig:3}b insert shows the decay of the measured storage efficiency. An exponential fit yields a coherence time of $2.6\pm0.3$ $\mu s$, consistent with our previous study \cite{wang2022} and physical parameters such as beam size and buffer gas pressure, suggesting that the limiting factor of the storage time is due to atomic diffusion. 

\subsubsection{Analysis}
In the first study, we normalize SNR to obtain $\rm{SNR}_{\langle n \rangle =1}$. In the memory source study, however, we argue that the directly measured SNR is what matters, because it represents the highest performance achievable of the two systems interfacing together. We show that these two cases are consistent with each other. For a WCS pulse, the mean photon number $\langle n\rangle$ in user adjustable. For an entanglement source, $\langle n\rangle$ is simply the heralding efficiency $\eta$ \cite{source2024}, defined as the ratio between the telecom-NIR pair rate and the telecom photon rate, i.e., $\langle n\rangle=\eta$. One can immediately infer that, if everything remains unchanged, the memory-source SNR is related to the memory-solo SNR via
\begin{equation}
    \label{eq:1}
\rm{SNR} \approx \eta \times \rm{SNR}_{\langle n \rangle =1} 
\end{equation}
Here with an optical cavity on the telecom photon path, a detection of a telecom photon projects the NIR photon onto a single spectral mode (see \ref{sec:app}) and results in $\eta=20(1)\%$. Using Eq. \ref{eq:1}, we would predict the SNR to be $\rm{SNR}_{\rm predict} = 95\times 0.2= 19(2)$. However, Eq. \ref{eq:1} assumes same memory performance, but we have worse storage efficiency in the memory-source case (5.2\%) than the memory-solo case (9.5\%), likely due to the bandwidth mismatch (see next paragraph). This additional factor would further lower the prediction to $\rm{SNR}_{\rm predict}=19\times5.2/9.5= 10(1)$, which agrees well with our experimental data.

A notable feature in Fig. \ref{fig:3} is that the bandwidths of the input pulses are larger than the output pulses. We take the Fourier transform of the pulse profiles in the time domain to get the frequency spectrum and obtain the FWHM bandwidth. For the memory-source results, we have 201(5) MHz for the input and 61(2) MHz for the output. We also measure the source solo bandwidth by bypassing the memory to get 266(5) MHz. It has been studied extensively both in our last work \cite{wang2022} and other groups \cite{Wei2020} that the bandwidth of EIT quantum memories scales with the control field $\Omega_c$, and that in optimal storage \cite{Gorshkov2007} the input and output pulses are time reversal of each other. In our case, however, we are limited by the available laser power $\Omega_c$ and can not reach the optimal storage regime. Therefore our retrieved pulses have smaller bandwidths than the input pulses. Additionally, the difference between storage efficiency in the memory-source case (5.2\%) versus the memory-solo case (9.5\%) is likely due to this bandwidth mismatch.

\section{high-fidelity entanglement}
\label{sec:entanglement}

\begin{figure}[t!]
\includegraphics[width=8.6cm]{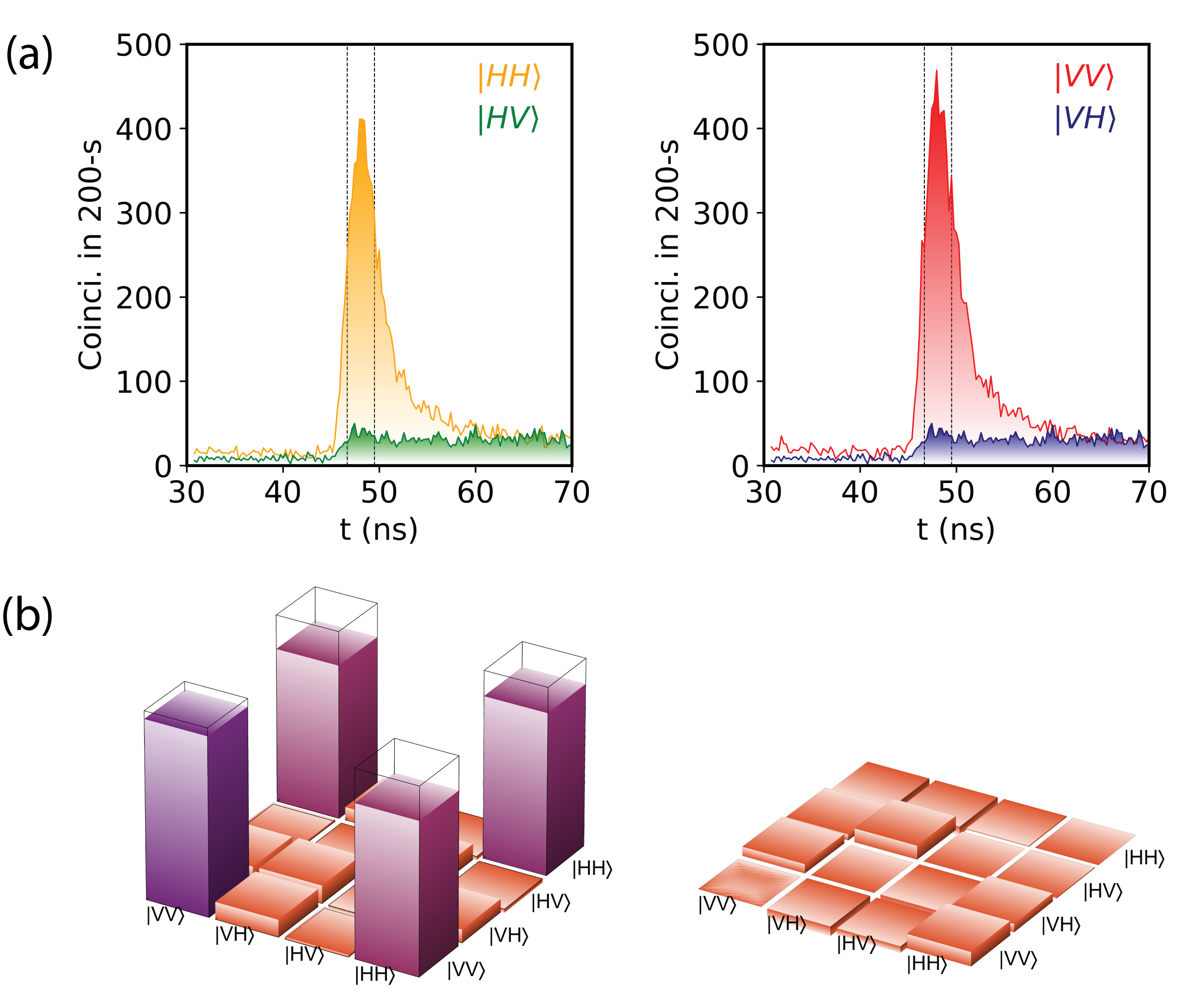}
\caption{Quantum state tomography result. (a) Exemplary traces of the retrieved photons measured in different polarization bases, showing both high entanglement $\mathcal{F}$ and low operation error (e.g., uniformity between two spatial rails). The dashed lines mark the region used for state tomography. (b) The reconstructed density matrix (left: real part; right: imaginary part) of the two-photon state puts a lower bound on the fidelity to be 86.5\%, with the expected state of $|\Phi^+\rangle=\frac{1}{\sqrt{2}}(|HH\rangle+|VV\rangle)$. The mesh grid (left) marks the expected state. 
}
\label{fig:4}
\end{figure}

Having demonstrated high-performance memory operation with single photons from the entanglement source, we move on to characterize the entanglement quality between the memory and the telecom photons,  
with the fidelity defined as $\mathcal{F}=\left(\textrm{Tr} \sqrt{\sqrt{\rho_0}\rho\sqrt{\rho_0}} \right)^2$, where $\rho_0=|\Phi^+\rangle\langle\Phi^+|$ and $\rho$ is the measured density matrix. We use two-photon quantum state tomography \cite{James2001} to reconstruct $\rho$. Specifically, we measure the cross-correlations between the telecom and NIR photons (retrieved after 50 ns) in 16 pre-chosen polarization configurations. For this measurement, we use 10 mW coupling field and 0.1 mW pumping field.

We plot in Fig. \ref{fig:4}a representative histograms zoomed on the retrieved photon, with data accumulated over 200 seconds and a time resolution of 256-ps.  
The high contrast between orthogonal bases and uniformity between $|HH\rangle$ and $|VV\rangle$ indicate a high degree of fidelity for the source-memory system. We use the same 2.82-ns detection window (dashed lines) for data analysis. 
Using a maximum likelihood method \cite{James2001}, the reconstructed density matrix is plotted in Fig. \ref{fig:4}b, with the mesh grid representing the ideal expected state $|\Phi^+\rangle$. This method establishes a lower bound on $\mathcal{F}$ at 86.5\%. 
This high entanglement fidelity between a telecom photon and a quantum memory represents a significant milestone, positioning the warm-vapor system among state-of-the-art physical systems \cite{Krutyanskiy2024,Leent2020,Tchebotareva2019,Jelena2022,Davidson2020,Chang2019}. To our knowledge, this is the first demonstration of entanglement between a warm-vapor system and a telecom photon.

To better understand what limits $\mathcal{F}$, we examine the SNR of the combined system. 
Assuming a noise model of the Werner state \cite{werner1989} in the form of $\rho=a|\Phi^+\rangle\langle \Phi^+| + \frac{(1-a)}{4}|I\rangle\langle I|$, where $a$ is related to SNR via $\rm{SNR}=\frac{2a}{1-a}$, we obtain a theoretic value of $\mathcal{F}$: 
\begin{equation}
    \label{eq:2}
    \mathcal{F_{\rm{theory}}}=1-\frac{3}{2 (\rm{SNR} +2)}
\end{equation} 

In the Sec. \ref{sec:memory}, we obtain a memory source SNR of 9.8(3). Using Eq. \ref{eq:2} we predict an $\mathcal{F_{\rm{theory}}}$=87(1)\%, in excellent agreement with the QST result in Fig.\ref{fig:4} (86.5\%). We therefore validate the SNR as an effective parameter to estimate $\mathcal{F}$.

\section{Utility time and entanglement generation rate}
\label{sec:extra}
\begin{figure}[t!]
\includegraphics[width=8.6cm]{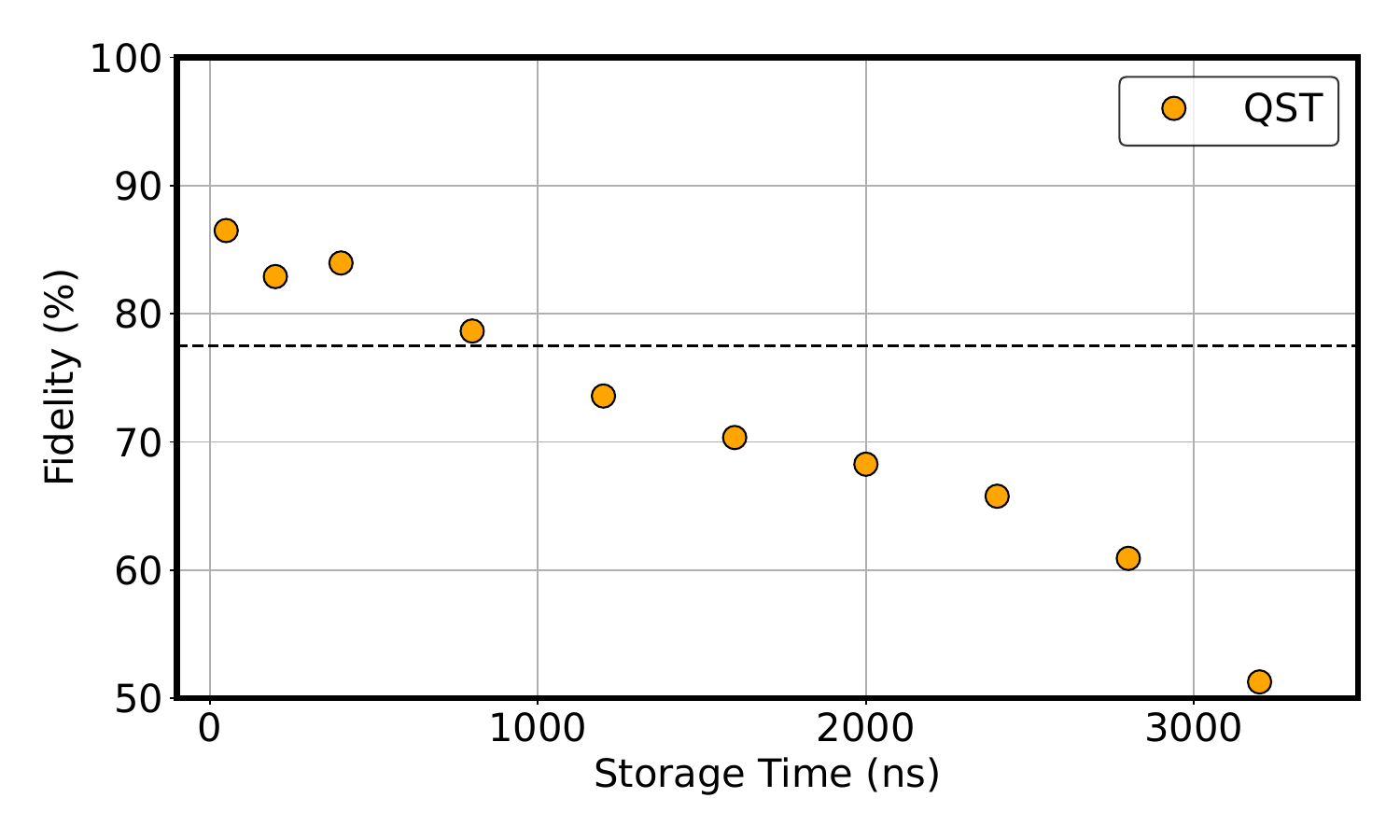}
\caption{Memory utility time. 
The orange circles represent the source-memory entanglement fidelity $\mathcal{F}$ based on QST as a function of the memory storage time. 
The dashed line marks $\mathcal{F}=77.5$\%, a threshold corresponding to $S=2$ in a CHSH measurement for violating the Bell inequality. Error bars represent one standard deviation. }
\label{fig:5}
\end{figure}
In this section, we perform additional analysis to highlight two important advantages of our system: utility time and high-rate entanglement generation, both of which are critical to efficiently distribute entanglement over long distances. 

First, we study how the entanglement fidelity $\mathcal{F}$ decays over time and plot the results in Fig. \ref{fig:5}. We define this as the ``utility time'' which provides more information than just the memory coherence time shown in Fig. \ref{fig:3}, because different protocols have different requirements on the lowest fidelity.  \cite{Neumann2021,Jelena2023,Bennett1996,Zhou2024,Arian2024}.
The dashed line marks 77.5\% which corresponds to a CHSH $S$ value of $S=2$ \cite{Clauser1969} (assuming a noise model of the Werner state \cite{werner1989}), which is required to achieve a entanglement based quantum security \cite{Neumann2021,Jelena2023}. For other protocols such as entanglement purification, a generally agreed threshold is 50\% \cite{Bennett1996,Zhou2024,Arian2024}. By these different thresholds, our current system possesses a utility time of $1\sim3$ $\mu s$. 
As briefly mention in Sec. \ref{sec:memory}, the coherence time is mostly limited by the atomic diffusion and consistent with our previous study \cite{wang2022}. This type of decoherence can be efficiently managed by adjusting the beam size and the buffer gas pressure to reach a long storage time of $\sim 1$ millisecond. However, extending the storage time while maintaining other performances (bandwidth, fidelity, etc) would require narrow-diameter vapor cells with anti-relaxation coated walls. We note very encouraging results \cite{Dideriksen2021} showing near-millisecond storage time with an alkane-coated cell. We highlight this storage time advantage which is critical to long-distance entanglement distribution, and our $\Lambda-$type memory (where photons are stored as ground state atomic excitations) offers a unique edge compared to ladder-type memories in warm vapors, where the storage time is limited to tens of nanoseconds due to the atomic excited lifetime \cite{Benjamin2024,FLAME,ORCA}.

\begin{figure}[t!]
\includegraphics[width=8.6cm]{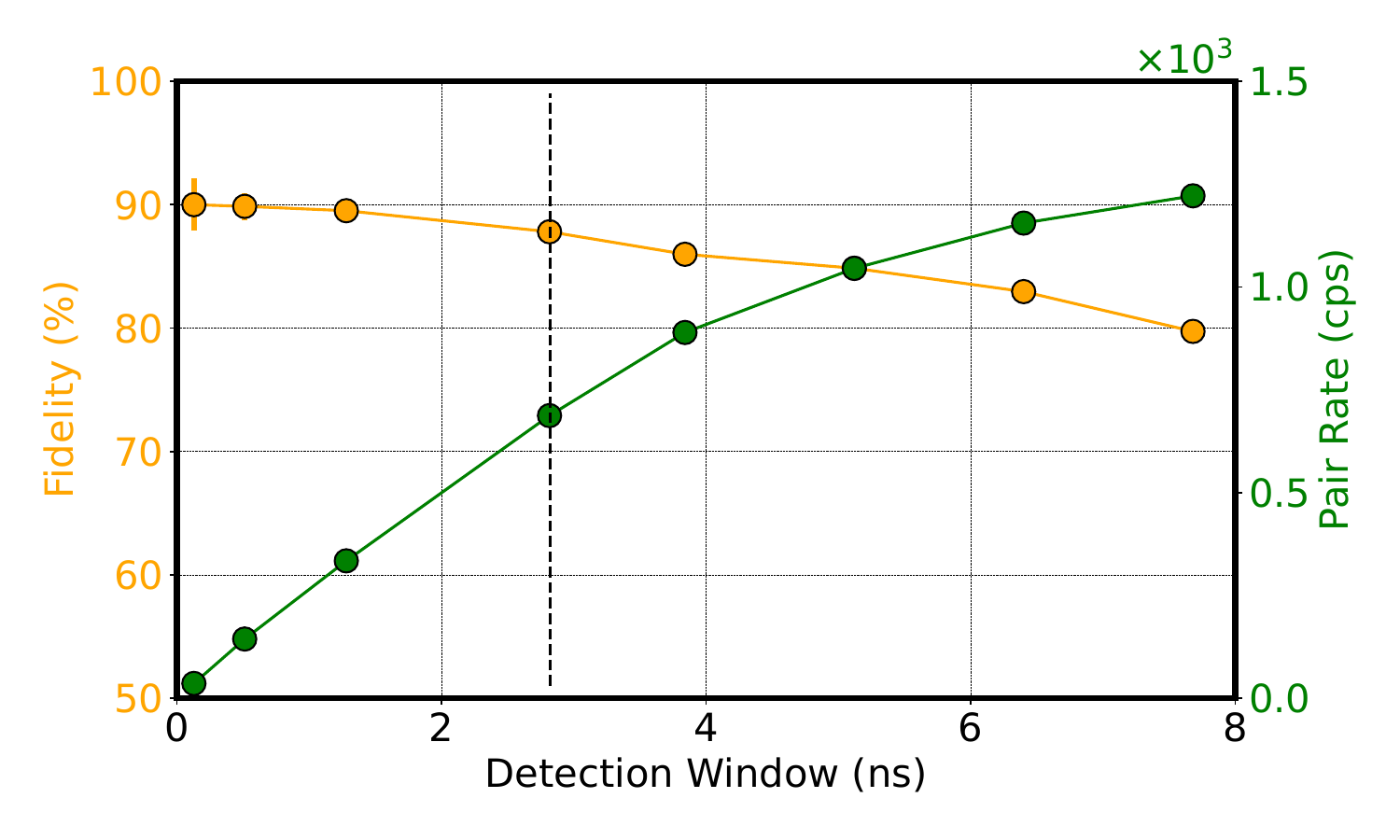}
\caption{High-rate and high-fidelity entanglement generation. We reanalyze data from Fig. \ref{fig:3}b to obtain the photon pair rates after the memory retrieval and the predicted fidelity as a function of the size of the detection window. Error bars represent one standard deviation, and are in some cases smaller than the data points.
}
\label{fig:6}
\end{figure}

Second, we calculate the entanglement generation rate. 
We reanalyze the data in Fig. \ref{fig:3}(b) and plot the rate and the predicted fidelity in Fig. \ref{fig:6}. Both of these parameters depend on the size of the detection window. To find the total number of entangled photon pairs coming out of the memory, we integrate over the retrieved pulse area and adjust for the following systematic factors: QST transmission (90\%), detector efficiency (65\%), and the undetected $|VV\rangle$ pair (50\%).  
We plot both the rate (orange circles) and $\mathcal{F_{\rm{theory}}}$ (green circle) as a function of the detection window size. The dashed line marks the 2.82-ns window. The trade-off between rate and fidelity is clear, with  $\mathcal{F_{\rm{theory}}}=90.2\%$ reachable at the cost of a lower rate. We achieve an entanglement generation rate of $1.2 \times 10^3$ pairs per second with an $ \mathcal{F_{\rm{theory}}}= 80 $\% for a detection window of 7.7 ns. This rate and fidelity are comparable to or better than existing demonstrations \cite{Zhang2024,Stephenson2020,Jelena2023}. The successful rate per trial,  $1.65(2)\times10^{-3}$, is also close to the state of the art \cite{Stephenson2020}. 

We attribute this high entanglement generation rate to the high duty cycle of the warm-vapor system, a unique advantage over other technologies. 
In our system, the overhead of preparing for memory operation---the downtime when the memory is unavailable for operation, is negligible: the only required step is the optical pumping that takes less than $1\mu s$ and is automatically applied at the end of the previous memory operation. This feature makes our memory essentially ever-ready. By comparison, in a cold-atom-based quantum memory, loading and cooling atoms takes $\sim$100 ms \cite{Hsiao2014}.

\section{conclusion and Outlook}
\label{sec:summary}

In summary, we have achieved high fidelity (up to 90.2\%) entanglement between a telecom photon and a room-temperature quantum memory, a crucial milestone towards large-scale quantum network. Utilizing FWM and EIT in Rb vapors with identical level structures, our source and memory work together natively, enabling entanglement generation of up to 1200 pairs per second at 80\% fidelity. 
Additionally, we show a memory with utility time of up to $3\mu s$, input bandwidth up to $2\pi \times266$ MHz, and source-independent performance $\rm{SNR}_{\langle n \rangle=1} = 95(5)$.

This work, coupled with our recent advancements in developing \cite{wang2022, source2024} and deploying \cite{APC2024} quantum networking hardware on telecommunication networks, establishes a foundation for practical applications such as field-deployed quantum repeaters, pending improvements in memory utility time. The current limitation stems solely from atomic diffusion, which can be addressed through anti-relaxation-coated vapor cells \cite{Dideriksen2021} while maintaining high bandwidth performance. 
Memory performance can be significantly enhanced by increasing laser power, potentially tripling bandwidth. Further improvements can be achieved through an optimized filter design that better matches bandwidth requirements while suppressing noise. 
These advancements are expected to yield significantly higher rates and fidelity, accelerating the transition of quantum networks from laboratory demonstrations to practical reality.

\section{ACKNOWLEDGMENTS} 
We thank the Qunnect Inc. team for their help with the manuscript. 
Special thanks to Felipe Giraldo and Niccol\`{o} Bigagli for their generous time helping with the manuscript structure throughout many scientific discussions.
\section{Appendix}
\label{sec:app}

\begin{figure}[]
\includegraphics[width=8.6cm]{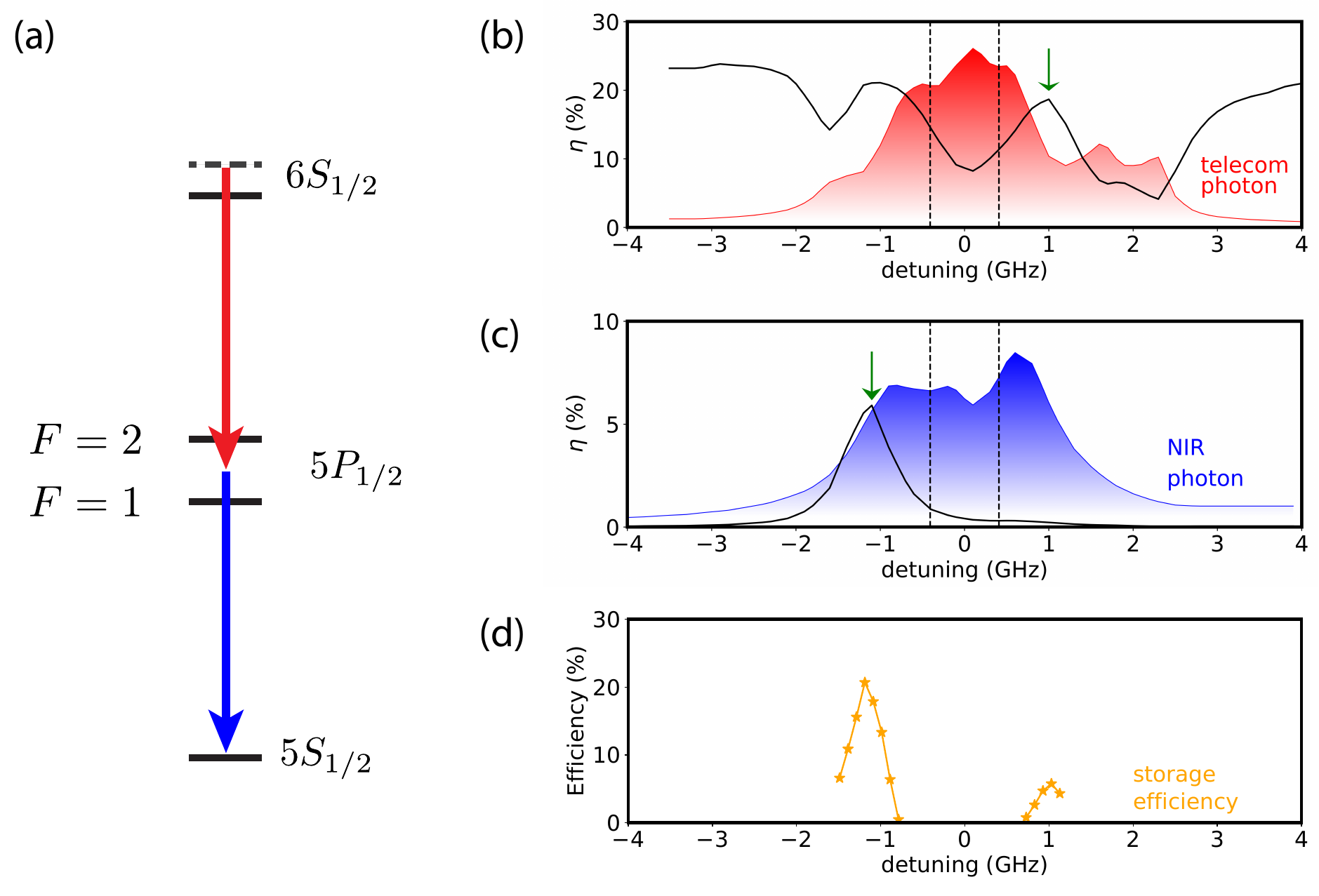}
\caption{Spectral mode matching. (a) Level scheme of the photon pair generation. There are two decay pathways due to the intermediate states $|5P_{1/2}, F=1,2\rangle$. 
(b) The telecom photon spectrum (red) in arbitrary unit and the heralding efficiency (solid line) obtained by scanning a telecom cavity (no cavity on the NIR photon). 
(c) The NIR photon spectrum (blue) in arbitrary unit and the heralding efficiency (solid line) obtained by scanning a NIR cavity (with telecom cavity fixed).
(d) The memory storage efficiency measured with a weak coherent pulse as a function of detunings.
}
\label{fig:7}
\end{figure}

Here, we discuss the details of spectral mode matching between the source and the memory. Our source works by off-resonant excitation that addresses atoms of different velocity classes \cite{source2024}. Hence the generated photons have a Doppler-broadened spectrum around the two decay pathways via the intermediate state $|5P_{1/2}, F=1,2\rangle$, see Fig. \ref{fig:7}a. 
Meanwhile, our memory, despite its high bandwidth, can only store NIR photons within a single spectral mode satisfying the two-photon resonance condition. 

Spectral mode-matching means that the NIR photons to be stored by the memory should always have the telecom photons entangled with them that triggers the memory operation. 
This metric can be conveniently evaluated by the heralding efficiency of the source, $\eta$, assuming no extra losses. Achieving high $\eta$ is critical to high entanglement fidelity because $\eta$ is proportional to the SNR of the combined system. 

To bridge the gap between the source and the memory, the simplest solution is to post-filter the telecom photon to a single spectral mode that is matched to the memory.  
Here, we use a monolithic optical cavity with an FWHM of $2\pi\times266$ MHz. 
We scan the cavity resonant frequency and measure the source photon properties (bypassing the memory). The results are shown in Fig. \ref{fig:7}b, where the red trace (in arbitrary units) represents the telecom photon rate, and the solid line represents $\eta$. The detuning is defined relative to the transition between the doubly excited state and the center of the $F=1,2$ states (marked by the two dashed lines). 

For large detunings (e.g., $|\delta| > 3$ GHz), $\eta$ is high but the rate is very low. For small detunings, we observe several major dips that one should avoid. We attribute the dip near 0 and -1.7 GHz to the absorption of the NIR photon by the ground state atoms ($^{87}$Rb and $^{85}$Rb) (the cell has $^{87}$Rb purity of 99\%), while the dip near 2 GHz to the on-resonantly generated photon due to $^{85}$Rb atoms. The final choice of operating point (near 1.1 GHz) is marked by the green arrow (see reasons below).

We then perform a scan to get the NIR photon spectrum with a NIR cavity (bandwidth = $2\pi\times397$ MHz) with the telecom cavity fixed at the operation point, see Fig. \ref{fig:7}c. Here the detuning is defined relative to the transition between $|5S_{1/2},F=2\rangle$ and the center of the $|5P_{1/2},F=1,2\rangle$ states. The small dip near $\delta\sim 0$ in the rate is expected due to the absorption from atoms in the ground state. Only a single peak in the heralding efficiency is observed because we have already selected a telecom mode. The lower $\eta$ is due to the frequency-independent loss through the memory and the NIR cavity. 
The green arrows mark the operating point that is -700 MHz detuned from the nearby transition $|5S_{1/2},F=2\rangle \rightarrow |5P_{1/2},F'=1\rangle$. 

Finally, we study the memory storage efficiency as a function of detuning, using a weak coherent pulse generated by a laser and pulse-shaped to optimize the efficiency \cite{Novikova2007}. Due to the destructive interference from the two hyperfine states $|5P_{1/2},F=1,2\rangle$ we obtain a memory efficiency with two peaks (Fig. \ref{fig:7}d). This behavior is also observed in other groups (\cite{Buser2022}). 
We note that those two peaks almost align with the two local maxima observed in (b). We proceed by selecting the one with the highest storage efficiency.

\bibliographystyle{apsrev4-2.bst}
\bibliography{library}

\end{document}